\documentstyle[a4,12pt]{article}

\begin{document}

\newcommand{\zb}{\bar z}
\newcommand{\th}{\theta}
\newcommand{\tb}{\bar{\theta}}
\newcommand{\zt}{\tilde z}
\newcommand{\zbt}{\tilde{\zb}}
\newcommand{\tht}{\tilde{\th}}
\newcommand{\tbt}{\tilde{\tb}}
\newcommand{\TH}{\Theta}
\newcommand{\TB}{\bar{\Theta}}

\newcommand{\la}{\lambda}
\newcommand{\LA}{\Lambda}
\newcommand{\DE}{\Delta}
\newcommand{\OM}{\Omega}
\newcommand{\SS}{{\bf S\Sigma}}

\newcommand{\pa}{\partial}
\newcommand{\pab}{ \bar{\partial} }
\newcommand{\dab}{\bar D}

\newcommand{\pat}{ \tilde{\pa} }
\newcommand{\pabt}{ \tilde{\pab} }
\newcommand{\dt}{ \tilde{D} }
\newcommand{\dbt}{ \tilde{\dab} }

\newcommand{\ZB}{\bar Z}
\newcommand{\HT}{ {H_{\th}}^z }
\newcommand{\HB}{ {H_{\tb}}^z }
\newcommand{\HO}{ H_{\th} ^{\ \th} }
\newcommand{\HZ}{ H_{\zb} ^{\ \th} }
\newcommand{\HZB}{ H_{\zb} ^{\ z} }
\newcommand{\HOB}{ H_{\tb} ^{\ \th} }

\newcommand{\KT}{ {k_{\tb}}^{\zb} }
\newcommand{\KB}{ {k_{\th}}^{\zb} }
\newcommand{\KO}{ k_{\tb} ^{\ \tb} }
\newcommand{\KZ}{ k_{z} ^{\ \tb} }
\newcommand{\KOB}{ k_{\th} ^{\ \tb} }
\newcommand{\KZB}{ k _{z}^{\ \zb} }
\newcommand{\LB}{\bar{L}}
\newcommand{\BT}{\bar{T}}
\newcommand{\HKB}{ \hat{k} _{\th}^{\ \zb} }
\newcommand{\HKOB}{ \hat{k} _{\th}^{\ \tb} }
\newcommand{\HKZB}{ \hat{k} _{z}^{\ \zb} }
\newcommand{\HKZ}{ \hat{k} _{z}^{\ \tb} }

\newcommand{\TT}{ {\cal T}_{\th z} }
\newcommand{\TZ}{ {\cal T}_{\tb \zb} }

\newcommand{\ro}{ {\stackrel{\circ}{\rho}} }
\newcommand{\RR}{ {\stackrel{\circ}{R}} }
\newcommand{\rr}{ {\stackrel{\circ}{r}} }
\newcommand{\cc}{ {\stackrel{\circ}{\chi}} }
\newcommand{\ee}{ {\stackrel{\circ}{\eta}} }
\newcommand{\GG}{ {\stackrel{\circ}{\Gamma}} }
\newcommand{\GAM}{ {\stackrel{\circ}{\gamma}}_{\TH} }
\newcommand{\gam}{ {\stackrel{\circ}{\gamma}}_{\th} }
\newcommand{\ga}{ {\stackrel{\circ}{\gamma}} }

\newcommand{\eb}{ \bar{\eta} }
\newcommand{\ebb}{ {\stackrel{\circ}{\eb}} }

\newcommand{\bfs}{{\bf \Sigma}}

\newcommand{\jb}{\bar{j}}
\newcommand{\tba}{\bar{t}}

\hfill MPI-Ph/92-78

\thispagestyle{empty}

\bigskip
\bigskip
\bigskip
\begin{center}
{\bf \Huge{On the Holomorphic Factorization}}
\end{center}
\begin{center}
{\bf \Huge{for Superconformal Fields}}
\end{center}
\bigskip
\bigskip
\bigskip
\bigskip
\bigskip
\bigskip
\centerline{\bf Fran\c cois Gieres$^{\, \dag \, \S}$}
\bigskip
\bigskip
\centerline{\it Max-Planck-Institut f\"ur Physik}
\centerline{\it Werner-Heisenberg-Institut}
\centerline{\it F\"ohringer Ring 6}
\centerline{\it D - 8000 - M\"unchen 40}
\bigskip
\bigskip
\bigskip
\bigskip
\bigskip
\bigskip
\bigskip
\bigskip
\bigskip
\bigskip
\bigskip
\bigskip
\bigskip

\begin{abstract}
\bigskip

For a generic value of the central charge,
we prove the holomorphic factorization
of partition functions for free superconformal fields
which are defined
on a compact Riemann surface without boundary.
The partition functions are viewed as functionals of the
Beltrami coefficients and their fermionic partners
which variables parametrize superconformal classes
of metrics.

\end{abstract}

\bigskip
\bigskip
\bigskip
\bigskip
\bigskip

\nopagebreak
\begin{flushleft}
\rule{2in}{0.03cm} \\

{\footnotesize \ ${}^{\dag}$
Alexander von Humboldt Fellow.}
\\  [-0.04cm]
{\footnotesize \ ${}^{\S}$
E-mail address: FRG@DM0MPI11.}
\\  [0.5cm]

MPI-Ph/92-78   \\
\vskip 0.07truecm
September 1992
\end{flushleft}

\newpage

\setcounter{page}{1}

\section{Introduction}

The fundamental theorem on holomorphic factorization in string theory
goes back to the work of
Belavin and Knizhnik \cite{bk} (see \cite{dph} and
\cite{kls} for further elaborations
and references):
this result concerns the factorization of partition functions
for free conformal fields on a compact Riemann surface without boundary.
Recently \cite{kls}, this theorem was extended from the case of
vanishing central charge to the general case, the partition functions being
regarded as functionals on the space of Beltrami coefficients parametrizing
complex structures on the Riemann surface. This extension relies on
a change of the renormalization prescription which amounts
to the addition
of a local counterterm shifting the Weyl anomaly to the chirally split
diffeomorphism anomaly.
Since the supersymmetric generalization of this local counterterm
was recently determined \cite{agn}, we can now address the problem
of holomorphic factorization (for arbitrary central charge)
in superstring theory along the lines of reference \cite{kls}.
For previous investigations with vanishing central charge,
we refer to \cite{dph} and the references given therein.

We work in component field formalism on a compact Riemann
surface $\bfs$ without boundary and we rely on the articles
\cite{dg} and \cite{agn}: in the first of these references,
the classification of
superconformal structures has been discussed and in the second, the
local counterterm that we use in the present work
was derived. Although all of these
results are based on the geometry of {\em super}
Riemann surfaces and although
our starting
equations are derived from superfield considerations, we
carry out the main computation in the
component field formalism, since some of the initial superspace
expressions are quite complex and not very transparent.

The space of superconformal structures
on the Riemann surface
$\bfs$ is parametrized by the Beltrami coefficient $\mu \equiv
\mu_{\zb} ^{\ z}$
and its fermionic partner, the Beltramino $\alpha \equiv
\alpha_{\zb}^{\ \th}$,
both variables
depending on a reference coordinate system ($z, \zb$).
The partition function $\Gamma$
is regarded as a functional on this space;
it is given by the properly renormalized superdeterminant
of the Laplacian associated to a system of free superconformal fields
(super bc system).
We are interested in proving the holomorphic factorization theorem
\begin{equation}
\label{1}
\delta_{\bar{\mu}, \bar{\alpha}} \;
\delta_{\mu, \alpha} \;
\left[ \, \Gamma \, + \, \Delta \Gamma \, \right] \ = \ 0
\ \ \ .
\end{equation}
Here,
$\Delta \Gamma$ is made up of local counterterms which allow to shift the
super Weyl anomaly to the chirally split form of the
superdiffeomorphism anomaly \cite{agn};
$\delta_{\mu , \alpha}$ ($\delta_{\bar{\mu}, \bar{\alpha}}$) represents
the variation w.r.t. $\mu$ and $\alpha$
($\bar{\mu}$ and $\bar{\alpha}$).

\section{Free superconformal fields}

Component field equations
on the Riemann surface ${\bf \Sigma}$
are best obtained from superfield expressions
defined on a super Riemann surface ${\bf S\Sigma}$ containing
${\bf \Sigma}$.
We now present the relevant superspace expressions
and then project them to component field equations.
We use the
notation and formalism of reference \cite{dg} with $\HT = 0$
(see also appendices
A and B of reference \cite{agn}).

The action for a super bc system of spin ${j \over 2}$ reads
\begin{equation}
\label{10}
S_j \ = \ \int_{{\bf S\Sigma}} d^4Z \  {\cal B}_{1-j} \, D_{\TB}
{\cal C}_{j} \ \equiv \
\int_{{\bf S\Sigma}} dZ \, d\ZB \, d\TH \, d\TB
\ {\cal B}_{1-j} \, D_{\TB} {\cal C}_{j}
\ \ ,
\end{equation}
where ${\cal B}$ and ${\cal C}$ transform according to
\begin{eqnarray}
{\cal B}_{1-j} ^{\prime}
( Z^{\prime} , \ZB^{\prime} ,\TH^{\prime} , \TB^{\prime} ) & = &
{\cal B}_{1-j} ( Z, \ZB ,\TH , \TB ) \ (D_{\TH} \TH^{\prime} )^{-(1-j)}
\nonumber  \\
{\cal C}_{j} ^{\prime}
( Z^{\prime} , \ZB^{\prime} ,\TH^{\prime} , \TB^{\prime} )  & = &
{\cal C}_{j} ( Z, \ZB ,\TH , \TB ) \ (D_{\TH} \TH^{\prime} )^{-j}
\end{eqnarray}
under a superconformal change of local coordinates
$( Z, \ZB ,\TH , \TB ) \to
( Z^{\prime} , \ZB^{\prime} ,\TH^{\prime} , \TB^{\prime} )$
on ${\bf S\Sigma}$.
If we redefine the variables ${\cal B}$ and ${\cal C}$ by virtue of the
integrating factor $\LA$ which depends on the reference coordinates
$( z, \zb ,\th , \tb )$,
\begin{equation}
{\cal B}_{1-j} ( Z , \ZB ,\TH , \TB ) \, \equiv \,
\left( B_{1-j} \, \LA^{-(1-j)/ 2} \right)
( z, \zb ,\th , \tb )
\quad , \quad
{\cal C}_{j}
( Z , \ZB ,\TH , \TB ) \, \equiv \,
\left( C_{j} \, \LA^{-j/2} \right) ( z, \zb ,\th , \tb )
\ \ ,
\end{equation}
we get the local action \cite{dg}
\begin{equation}
\label{13}
S_j \ = \ \int_{{\bf S\Sigma}} d^4z \;
{1 \over (\KO)^2 } \; B_{1-j}
\left[  \, ( \dbt - \KT \pabt ) - {j \over 2} \, (\pa \HB - \KT \pa \HZB )
\, \right] C_{j}
\ \ .
\end{equation}
This functional of $B$ and $C$
does not depend on $\LA$, but only on the Beltrami superfield
$\HB$ which contains the ordinary Beltrami coefficient $\mu$ and its fermionic
partner $\alpha$ among its components: by substituting
the $\th$-expansions (of the WZ-gauge)
\begin{equation}
\HB \, = \, \tb \, \mu + \th \tb \, [ -i \alpha ]
\quad , \quad
C_{2j} \, = \, c_{j} + \th \, [ i \epsilon_{j+{1 \over 2}} ]
\quad , \quad
B_{1-2j} \, = \, i \beta_{{1 \over 2}-j} + \th \, b_{1-j}
\ \ ,
\label{14}
\end{equation}
we get the component field action
\begin{eqnarray}
\label{15}
S_{2j} & = & - \int_{{\bf \Sigma}} d^2z \;
\left\{ b \left[ \, [ \pab - \mu \pa - j (\pa \mu ) ] c
+ {1\over 2} \, \alpha \epsilon \, \right] \right.
\nonumber \\
& & \left.
\qquad \qquad
+ \beta
\left[ \, [ \pab - \mu \pa - (j + {1\over 2} ) (\pa \mu ) ] \epsilon
+ {1\over 2} \, \alpha (\pa c) - j c (\pa \alpha ) \, \right] \right\}
\ \ .
\end{eqnarray}
Here, $b,c,\beta, \epsilon$ are ordinary conformal fields and
we recognize the action for an ordinary bc system of spin $j$.

Following reference \cite{kls}, we now introduce a supermetric \cite{agn}
\begin{equation}
\label{16}
\ro_{Z\ZB} \ \equiv \ \LA^{-1} \, \bar{\LA}^{-1} \, \ro_{z\zb}
\ \ .
\end{equation}
Furthermore, for a superconformal field ${\cal C}_{j,\jb}
( Z, \ZB ,\TH , \TB )$ transforming as
$$
{\cal C}_{j,\jb}^{\prime} \ = \
{\cal C}_{j,\jb} \; (D_{\TH} \TH^{\prime} )^{-j}  \;
(D_{\TB} \TB^{\prime} )^{-\jb}
$$
and related to $C_{j,\jb} (z,\zb , \th ,\tb)$ by
\[
{\cal C}_{j,\jb} \ \equiv \
C_{j,\jb} \  \LA^{-j / 2} \
\bar{\LA} ^{-\jb / 2} \,
\ \ ,
\]
we define a norm
\begin{eqnarray}
\mid \mid {\cal C}_{j,\jb} \mid \mid ^2 & \equiv &
\int_{{\bf S\Sigma}} d^4Z \;
\mid {\cal C}_{j,\jb} \mid ^2 \ ( \ro_{Z\ZB} )
^{{1 \over 2} (1 -j-\jb)}
\nonumber  \\
& = &
\int_{{\bf S\Sigma}} d^4z \;
{\sqrt{A} \over \KO} \
\mid  C_{j,\jb} \mid ^2 \ ( \ro_{z\zb} ) ^{{1 \over 2} (1 -j-\jb)}
\ \equiv \
\; \mid \mid C_{j,\jb} \mid \mid ^2
\ \ .
\end{eqnarray}
Here, $A$ and $\KO$ are polynomials in $\HB$ and its complex conjugate and
in their derivatives.
For instance, $\mid \mid D_{\TB} {\cal X} \mid \mid ^2$
represents the standard action for a scalar superfield ${\cal X}$
for which functional
the component field expression can be found in
\cite{dg}. For ${\cal C}_{j,\jb} = D_{\TB} {\cal C}_j$, it is easy to
derive an explicit expression for the norm using appendix A of
reference \cite{agn}: from the $\th$-expansions (WZ-gauge)
of $\HB, \, C_{2j}$ and
\begin{equation}
\label{47}
\ro_{z\zb} \ = \ \ro_0 \, + \, \th \, [ i \ro_1] \, + \, \tb \, [-i
\stackrel{\circ}{\bar{\rho}} _1 ] \, + \, \th \tb \, [- {1\over 2} \,
\ro_1 \stackrel{\circ}{\bar{\rho}} _1  / \ro_0 ]
\ \ ,
\end{equation}
one then finds
\begin{equation}
\mid \mid D_{\TB} {\cal C}_{2j} \mid \mid ^2 \; = \;
\int_{{\bf \Sigma}} d^2z \ {1\over 1 -\mu \bar{\mu}} \
( \ro_0 ) ^{-j} \
\mid
[ \pab - \mu \pa - j (\pa \mu ) ] c
+ {1\over 2} \, \alpha \epsilon \mid ^2
\ \ .
\end{equation}
This expression encompasses the results of the bosonic theory \cite{kls}.

The super Laplacian $\DE_{2j}$ (acting on $C_{2j}$)
associated to the supermetric $\ro_{z\zb}$ is now defined by
\begin{equation}
\mid \mid D_{\TB} {\cal C}_{2j} \mid \mid ^2 \; = \;
< C_{2j} \! \mid \, - \DE_{2j} C_{2j} >
\end{equation}
and the partition function $\Gamma$ is given by the superdeterminant
of $\DE_{2j}$:
\begin{eqnarray}
\Gamma & \equiv & \Gamma [ \ro_0 ; \mu , \bar{\mu} , \alpha , \bar{\alpha} ]
\, - \,
\Gamma [ \ro_0 ; 0, 0 , 0, 0 ]
\label{39}  \\
\Gamma [ \ro_0 ; \mu , \bar{\mu} , \alpha , \bar{\alpha} ] & \equiv &
- {1 \over 2 } \ {\rm ln \; sdet} \,  ( -\DE_{2j} )
\ \ .
\nonumber
\end{eqnarray}
The superdeterminant
is assumed to be renormalized by the
$\zeta$-function and to involve a mass term accounting for global
zero modes \cite{kls}.

\section{The local counterterm $\DE \Gamma$}

By virtue of eq.(\ref{47}), the supermetric $\ro_{z\zb}$ has bosonic
(fermionic) components $\ro_0$ ($\ro_1$).
These variables give rise to a non-holomorphic
superaffine connection with components \cite{agn}
\begin{equation}
\label{60}
\GG_{z} \; = \;  \pa \, {\rm ln} \, \ro_0
\ \ \ \ \ \ , \ \ \ \ \ \
\ee_{\th} \; = \;  \ro_1 \, / \, \ro_0
\ \ \ .
\end{equation}
The corresponding 'field strengths' define a non-holomorphic
superprojective connection,
\begin{equation}
\label{61}
\rr_{zz} \; = \; ( \pa \GG_z - \frac{1}{2} \, \GG_z ^2 )
\, - \, \frac{1}{2} \, \ee_{\th} \  \pa \ee_{\th}
\ \ \ \ \ \ , \ \ \ \ \ \
\cc_{z\th} \; = \; (\pa - \frac{1}{2} \, \GG_z ) \, \ee_{\th}
\ \ \ ,
\end{equation}
while the 'field strength' of $\mu$ is given by the
supercovariant derivative
\begin{equation}
\label{62}
{\cal D} \mu \ = \
(\pa + \GG_z ) \, \mu
\, + \, \frac{1}{2} \, \ee_{\th} \, \alpha
\ \ \ .
\end{equation}
In order to simplify the notation in the following,
we will suppress the indices on all component fields.

The local counterterm relating the super Weyl and the chirally split
superdiffeomorphism anomalies involves the local functional \cite{agn}
\begin{eqnarray}
\Delta \Gamma & = &
\Gamma_{II} \, + \,
\Gamma_{III}
\nonumber
\\
\Gamma_{II} & = &
k \,
\int_{{\bf \Sigma}} d^2z \
\left\{  \, \mu  \, ( r - \rr ) \, - \, \alpha \, ( \chi - \cc ) \, \right\}
\ + \ {\rm c.c.}
\label{2}
\\
\Gamma_{III} & = &
-k \,
\int_{{\bf \Sigma}} d^2z \ \,
\frac{1}{1 - \mu \bar{\mu}} \ \left\{ \,
({\cal D} \mu ) \, ( \bar{\cal D} \bar{\mu} )
\; - \; \frac{1}{2} \,
\bar{\mu} \, ( {\cal D} \mu )^2
\; - \; \frac{1}{2} \,
\mu \, ( \bar{\cal D} \bar{\mu} )^2
 \, \right\}
\ \ \ .
\nonumber
\end{eqnarray}
Here,
$r \equiv r_{zz}(z)$ and $\chi \equiv \chi_{z\th}(z)$ denote, respectively, the
bosonic and fermionic components of a holomorphic
superprojective connection and
$$
k \ = \ {Q_j \over 12 \pi}
\qquad {\rm with} \quad Q_j = 6j (j-1) +1
$$
represents the central charge of our model.

\section{Variation of $\Delta \Gamma$ and $\Gamma$}

Since $\mu$ and $\alpha$ parametrize an infinite dimensional space ${\cal M}$,
it is preferable to replace the variation
$\delta_{\mu, \alpha}$ by
$d_t = dt \, \pa_t$ where $t$ is a local coordinate
labeling a finite dimensional analytic
family $(\mu_t , \alpha_t) \in {\cal M}_t
\subset {\cal M}$. From the expression (\ref{2}), we then conclude that
\begin{equation}
\label{50}
d_{\bar{t}} \, d_t \, \Delta \Gamma \ = \
k \,
\int_{{\bf \Sigma_t}} d^2z \ \,
\frac{1}{1 - \mu \bar{\mu}} \
\left[ \, (\pa + g ) \, d_t \mu
\; + \; \frac{1}{2} \,
\ee \, d_t \alpha \, \right] \,
\left[ \, (\pab + \bar{g} ) \, d_{\bar{t}} \bar{\mu}
\; + \; \frac{1}{2} \,
\ebb \,
d_{\bar{t}} \bar{\alpha} \, \right]
\ ,
\end{equation}
where
\begin{equation}
\label{51}
g \ \equiv  \ \GG \; - \; \frac{1}{1- \mu \bar{\mu}} \, \left[ \,
\bar{\cal D} \bar{\mu} \, - \, \bar{\mu}
\, {\cal D} \mu \, \right]
\end{equation}
and where ${\bf \Sigma}_t$ corresponds to a fibre of the bundle
${\bf \Sigma} \times {\cal M}_t$.

The variation of $\Gamma$
is given by the supersymmetric extension of the
index theorem for families of elliptic operators (see \cite{kls}
and references therein).
The local bosonic version of this theorem reads
\cite{kls}
\begin{equation}
\label{40}
d_{\tba} \, d_t \, \Gamma_{bosonic} \; = \;
k \, \int_{{\bf \Sigma_t}} d^2 Z \, d\tba \, dt \mid \pa_t
\pa_{\ZB} \, {\rm ln} \, \ro_{Z\ZB} (Z, \ZB ) \mid ^2
\end{equation}
and its manifestly supersymmetric generalization is given by
\begin{equation}
\label{41}
d_{\tba} \, d_t \, \Gamma \; = \;
k \, \int_{{\bf S\Sigma_t}} d^4 Z \, d\tba \, dt \mid
\pa_t D_{\TB} \, {\rm ln} \, \ro_{Z\ZB}  \mid ^2
\ \ .
\end{equation}
Here, $\ro_{Z \ZB}
(Z , \ZB , \TH , \TB )$ is the supermetric introduced in section 2.
An explicit expression for $D_{\TB} \, {\rm ln} \, \ro_{Z\ZB}$
in terms of the superfields $\HB$ and $\LA$ is given
by the (complex conjugate of) eq.(30) of reference \cite{agn}.
This expression is quite complicated, but by going over to the reference
coordinates $( z, \zb , \th , \tb )$ in the integral (\ref{41})
and projecting to components, we find the simple result
\begin{equation}
\label{42}
k \, \int_{{\bf S\Sigma_t}} d^4 Z \, d\tba \, dt \mid
\pa_t D_{\TB} \, {\rm ln} \, \ro_{Z\ZB}  \mid ^2
\; = \;
-k \, \int_{{\bf \Sigma_t}} d^2 z \  \; {1 \over 1-\mu \bar{\mu}} \; \mid
(\pa + g ) \, d_t \mu
\; + \; \frac{1}{2} \, \ee \, d_t \alpha \mid ^2
\ \ .
\end{equation}
Obviously, the
right hand sides of eqs.(\ref{50}) and (\ref{42}) coincide
with each other up to an overall sign.
Combining these equations with the relation
(\ref{41}), we get the final result
\begin{equation}
d_{\tba} \, d_t \, \left[ \, \Gamma  + \DE \Gamma \, \right] \ = 0
\ \ ,
\end{equation}
i.e. the holomorphic factorization of $\Gamma  + \DE \Gamma$.

\section{Conclusion}

For a generic value of the central charge, we have shown that the
holomorphic splitting property in superstring theory can be
recovered on the space of superconformal structures
(parametrized by $\mu$ and $\alpha$) by the inclusion of the local
counterterm $\DE \Gamma$.

The holomorphic factorization for ordinary free conformal fields
has recently been generalized by including a coupling
to Yang-Mills connections on the Riemann surface \cite{kls2}.
By virtue of the results derived
in the present work, it should be possible to treat the
coupling of superconformal and super Yang-Mills
fields along the lines of reference \cite{kls2}.

\vskip 1.5truecm
{\bf \Large{Acknowledgements}}
\vspace {5mm}

It is a pleasure
to thank M.Knecht and J.P.Ader
for a critical reading of the manuscript.

\end{document}